**AoI-Guided Client Selection for Robust and Timely Federated Intrusion Detection in Cloud-Edge Security Analytics**

Chun Yin Chiu

King's College London

Abstract. Federated learning (FL) is attractive for cloud-edge intrusion detection because it enables collaborative training over distributed telemetry without centralizing raw logs. In production security analytics pipelines, however, only a subset of clients participates in each round, and heterogeneous bandwidth, stragglers, and dropouts can cause the server to rely on stale client information. This paper studies client participation as a timeliness-aware systems problem using Age of Information (AoI). We compare three lightweight policies for federated intrusion detection: AoI-first, utility-first, and a hybrid AoI+utility rule with a tunable trade-off parameter. Across CIC-IDS2017, NSL-KDD, ToN-IoT, and a synthetic drift benchmark under clean, poisoning, and poisoning-plus-robust-aggregation settings, AoI-aware selection reduces average AoI by about 39-41% and peak AoI by about 70% relative to random sampling while keeping the per-round communication budget fixed. The hybrid policy usually preserves or improves Macro-F1/AUC and provides an interpretable knob for balancing freshness, detection quality, and robustness. The results should be interpreted as a scheduling and timeliness study rather than a claim of uniformly improved detection accuracy: in some adversarial settings, freshness improvements trade off against false-positive rate. The main practical message is that cloud-scale, privacy-preserving intrusion analytics can improve timeliness through a lightweight scheduling layer without changing the underlying FL participation budget.

## Introduction

Modern enterprise monitoring is increasingly built on cloud and edge infrastructures that collect high-volume network telemetry, flow records, and host logs from geographically distributed sites. Training intrusion detection models by centralizing these logs is often undesirable because of privacy constraints, compliance requirements, storage overhead, and cross-domain data ownership. Federated learning (FL) provides a natural cloud-scale alternative: clients train local models close to where telemetry is generated, while a coordinating server aggregates only model updates. Yet in realistic deployments, only a subset of clients can participate in each round, and the platform must cope with heterogeneous bandwidth, stragglers, and dropouts. As a result, the global detector may rely on stale client information and react too slowly to evolving attacks.

We study this systems problem through Age of Information (AoI), a freshness metric that quantifies how outdated the server's information is for each client. Although AoI has been investigated in wireless and edge FL scheduling, its role in cloud-oriented intrusion analytics remains underexplored. This matters because FL-based IDS pipelines operate over dynamic traffic distributions: attack campaigns, workload mixes, and benign behavior all shift over time. When client participation is stale, the global detector can lag behind current conditions, reducing timely detection and sometimes increasing false alarms.

Our objective is to make federated IDS better aligned with cloud and big-data platforms by treating timeliness as a first-class systems objective alongside accuracy and robustness. To that end, we build an FL-IDS evaluation framework that reports Avg/Peak AoI together with Macro-F1, AUC, FPR, and communication cost, and we compare three simple selection policies under both clean and adversarial conditions. Across four datasets and three threat settings, AoI-aware scheduling materially improves freshness without increasing the standard per-round communication budget, while the hybrid policy is usually the best operating point.

The contributions are threefold. (1) We formulate client participation for FL-based intrusion detection as a timeliness-aware scheduling problem and argue that stale participation is an overlooked bottleneck for cloud-edge security analytics. (2) We propose lightweight AoI-guided selection policies that are easy to implement in existing FL control planes and preserve the usual communication budget. (3) We provide a focused multi-dataset empirical study across clean and poisoned settings, reporting accuracy-timeliness-robustness trade-offs and identifying the hybrid AoI+utility policy as a practical operating point rather than a universally dominant detector.

## Related Work

Federated learning was introduced for decentralized training with reduced communication (FedAvg) [1], and later extended to better handle heterogeneity and drift, e.g., FedProx [2] and SCAFFOLD [3]. Client selection has also been used to improve efficiency and convergence; Oort [4] is a prominent utility-guided scheduler for cross-device FL. In parallel, AoI has become a standard metric for information freshness [5], and recent wireless/edge FL works have shown that AoI-aware scheduling can reduce staleness [13-15]. For cybersecurity, FL-based IDS has been studied as a privacy-preserving alternative to centralized detection, with recurring concerns about non-IID traffic, communication constraints, and adversarial robustness [16]. Our contribution differs in emphasis: rather than treating FL-IDS solely as an accuracy problem, we frame client participation as a cloud-systems scheduling problem and evaluate whether a simple AoI signal can improve the operational freshness of distributed security analytics without increasing the communication budget.

## AoI-Aware Client Selection for Cloud-Edge FL-IDS

We consider a synchronous FL setting with $N$ clients. In each round $t$, the server selects $k$ clients to train locally and send model updates. Let $s_i$ denote the most recent round in which client $i$ successfully contributed an update. The client AoI at round $t$ is:

$$A_i(t) = t - s_i(t).$$

A higher AoI means the server's view of that client's data is more stale. We compare the following selection rules ($k$ clients per round):

- Random: Uniformly sample $k$ clients without replacement (baseline for FedAvg/FedProx/SCAFFOLD).
- AoI-first: Select the top-$k$ clients with the highest AoI (freshness-first).
- Utility-first: Select the top-$k$ clients with the highest utility score, inspired by Oort [4]. We estimate per-round utility from local loss, uplink bandwidth, and dropout probability, excluding AoI so that freshness can be studied separately. Concretely, we use a score proportional to $loss_i * \sqrt{bw_i} * (1 - dropout_i)$, favoring high-loss clients that are likely to return updates quickly. Here $loss_i$ is a client-side scalar estimated using the current global model on a small local minibatch (or local validation split); no raw data are shared. We smooth this signal with an exponential moving average to reduce noise before ranking.
- AoI+Utility: Min-max normalize AoI and utility to [0,1] within a round and rank clients by $score_i = \lambda * AoI_i + (1 - \lambda) * u_i$, where $\lambda$ controls the freshness-performance trade-off.

Aggregation. Unless otherwise specified, the server aggregates client models by weighted averaging (FedAvg) [1]. To study robustness, we additionally evaluate a coordinate-wise trimmed-mean aggregator [11], which removes extreme client updates before averaging, mitigating Byzantine/poisoning behavior [10-12].

## Experimental Setup for Cloud-Scale Federated IDS

Datasets. We evaluate on CIC-IDS2017 flows (Friday DDoS + PortScan mini subset) [6], NSL-KDD (5-class variant) [7], ToN-IoT network traffic (10 classes) [8,9], and a synthetic drift dataset designed to stress freshness under distribution shift. The CIC-IDS2017 experiments should be interpreted as reduced-subset results rather than full-dataset CIC-IDS2017 benchmark results. After preprocessing, NSL-KDD has 125,973 train / 22,544 test samples with 122 features; ToN-IoT has 105,521 train / 105,522 test samples with 118 features; synthetic drift has 150,000 train / 5,000 test samples with 20 features. These datasets collectively represent distributed security telemetry workloads that are natural candidates for privacy-preserving cloud or cloud-edge deployment.

FL configuration. Unless stated otherwise, we use $N=50$ clients and $T=60$ rounds, selecting $k=10$ clients per round (20% participation). Training data are partitioned non-IID using a Dirichlet split ($\alpha=0.3$). Each client performs 20 local SGD steps per round (batch size 128, learning rate 0.2) using a lightweight softmax classifier to isolate the effect of selection. To model system heterogeneity, each client is assigned a random uplink bandwidth bw_i ∈ [5,50] Mbps and a dropout probability p_i ∈ [0,0.25]; dropped clients do not contribute that round. We keep k fixed (20% participation) to isolate the effect of selection on timeliness and detection quality; therefore total communication is near-constant across selection policies.

Threat model and scenarios. We run three scenarios per dataset: (i) clean (no attack), (ii) poison, where a fraction rho=0.2 of clients are malicious and perform label-flip poisoning by flipping labels with probability 0.3 toward a fixed target class, and (iii) poison_robust, which combines the same poisoning with trimmed-mean aggregation (trim ratio 0.2) [11].

Metrics. Accuracy is measured by Macro-F1 and AUC (macro one-vs-rest for multi-class; binary AUC for binary labels when applicable), and we also report a benign-vs-attack false positive rate (FPR). Timeliness is measured by average AoI and peak AoI (in rounds). Communication cost is measured as total bytes exchanged (download+upload of model parameters) aggregated to MB.

## Results

We report last-round performance as mean +/- standard deviation over three random seeds. Table 1 summarizes the clean setting under FedAvg, highlighting random sampling versus AoI-aware policies (AoI-first and AoI+utility). Utility-first selection and additional algorithmic baselines (FedProx and SCAFFOLD) are shown in Figures 1-2 for context. To highlight robustness, Table 2 reports results under poisoning with trimmed-mean aggregation, comparing random sampling against guided selection. The purpose of the evaluation is to characterize scheduling trade-offs under a fixed participation budget, not to claim that AoI scheduling uniformly improves every detection metric.

*Table 1: Clean setting (FedAvg, last round). Macro-F1 and Avg AoI are mean+/-std over 3 seeds; other metrics are means.*

| Dataset | Method | Macro-F1 | AUC | FPR | Avg AoI | Peak AoI | Comm (MB) |
|---|---|---|---|---|---|---|---|
| CIC-IDS2017 | FedAvg-Rand | 0.654±.003 | — | 0.051 | 6.12±0.17 | 29.00 | 2.27 |
| CIC-IDS2017 | FedAvg-AoI | 0.655±.004 | — | 0.045 | 3.32±0.13 | 6.67 | 2.30 |
| CIC-IDS2017 | FedAvg-AoI+Util | 0.658±.004 | — | 0.028 | 3.39±0.05 | 7.33 | 2.29 |
| NSL-KDD | FedAvg-Rand | 0.482±.013 | 0.847 | 0.063 | 5.36±0.21 | 19.33 | 2.67 |
| NSL-KDD | FedAvg-AoI | 0.473±.020 | 0.824 | 0.063 | 3.42±0.16 | 6.67 | 2.63 |
| NSL-KDD | FedAvg-AoI+Util | 0.492±.019 | 0.855 | 0.066 | 3.52±0.14 | 7.00 | 2.61 |
| ToN-IoT | FedAvg-Rand | 0.639±.015 | 0.961 | 0.167 | 5.67±0.87 | 24.67 | 4.99 |
| ToN-IoT | FedAvg-AoI | 0.621±.031 | 0.964 | 0.064 | 3.42±0.18 | 6.67 | 4.92 |
| ToN-IoT | FedAvg-AoI+Util | 0.643±.025 | 0.963 | 0.146 | 3.69±0.17 | 7.67 | 4.90 |
| Synthetic-Drift | FedAvg-Rand | 0.171±.010 | 0.518 | 0.430 | 5.95±0.52 | 26.67 | 0.45 |
| Synthetic-Drift | FedAvg-AoI | 0.182±.006 | 0.521 | 0.434 | 3.37±0.08 | 7.00 | 0.45 |
| Synthetic-Drift | FedAvg-AoI+Util | 0.187±.010 | 0.520 | 0.268 | 3.37±0.13 | 6.67 | 0.45 |

*Fig. 1 shows the accuracy-timeliness trade-off on CIC-IDS2017 (clean). Random sampling yields reasonable Macro-F1 but high staleness (Avg AoI about 5.8 rounds; Peak AoI about 29 rounds). AoI-first and AoI+utility move the operating point left substantially (about 40-45% lower Avg AoI and about 70-75% lower Peak AoI) with comparable Macro-F1, indicating that prioritizing stale clients improves freshness without necessarily sacrificing detection quality. Utility-first can emphasize performance but tends to increase staleness, motivating the hybrid policy.*

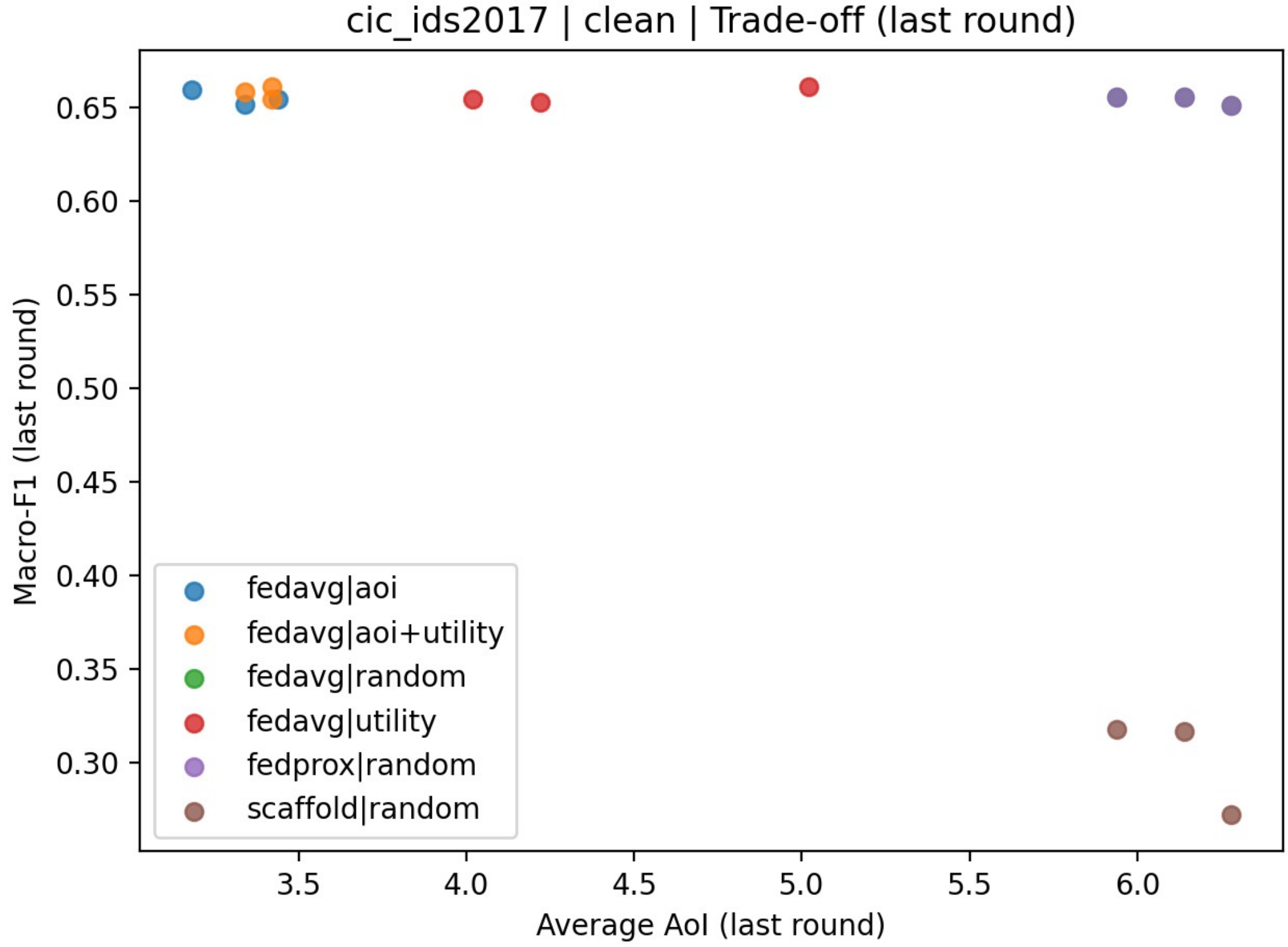

Fig. 1. Accuracy-timeliness trade-off on CIC-IDS2017 mini subset (clean). Points show mean last-round Macro-F1 versus Avg AoI over 3 seeds.

*Table 2: Poisoning + trimmed-mean robust aggregation (FedAvg variants, last round). Macro-F1 and Avg AoI are mean+/-std over 3 seeds; other metrics are means.*

| Dataset | Method | Macro-F1 | AUC | FPR | Avg AoI | Peak AoI | Comm (MB) |
|---|---|---|---|---|---|---|---|
| CIC-IDS2017 | FedAvg-Rand | 0.648±.002 | — | 0.086 | 5.94±0.99 | 27.67 | 2.27 |
| CIC-IDS2017 | FedAvg-Util | 0.656±.007 | — | 0.039 | 4.54±1.03 | 19.00 | 2.27 |
| CIC-IDS2017 | FedAvg-AoI+Util | 0.649±.002 | — | 0.081 | 3.38±0.08 | 6.67 | 2.29 |
| NSL-KDD | FedAvg-Rand | 0.505±.008 | 0.904 | 0.069 | 5.74±0.62 | 23.67 | 2.60 |
| NSL-KDD | FedAvg-Util | 0.533±.036 | 0.900 | 0.074 | 5.10±0.76 | 18.67 | 2.59 |
| NSL-KDD | FedAvg-AoI+Util | 0.514±.005 | 0.905 | 0.071 | 3.43±0.32 | 6.67 | 2.57 |
| ToN-IoT | FedAvg-Rand | 0.627±.012 | 0.961 | 0.163 | 5.55±0.37 | 19.33 | 4.98 |
| ToN-IoT | FedAvg-Util | 0.628±.020 | 0.958 | 0.164 | 4.51±0.47 | 15.67 | 4.92 |
| ToN-IoT | FedAvg-AoI+Util | 0.635±.005 | 0.961 | 0.221 | 3.50±0.28 | 7.67 | 4.91 |
| Synthetic-Drift | FedAvg-Rand | 0.169±.021 | 0.521 | 0.634 | 5.70±0.30 | 26.67 | 0.45 |
| Synthetic-Drift | FedAvg-Util | 0.168±.006 | 0.528 | 0.817 | 3.62±0.12 | 9.33 | 0.44 |
| Synthetic-Drift | FedAvg-AoI+Util | 0.179±.010 | 0.523 | 0.551 | 3.31±0.20 | 6.33 | 0.45 |

Robustness under poisoning. Table 2 and Fig. 2 summarize the poison_robust setting. Trimmed-mean aggregation improves resilience to malicious label-flip clients [11,12], while selection still controls timeliness. Compared with FedAvg-Rand, the AoI+utility policy reduces Avg AoI by about 37-43% across datasets and generally matches or improves Macro-F1/AUC. We also report FPR; in some adversarial configurations, emphasizing freshness can increase false positives (for example, ToN-IoT under poison_robust), highlighting an accuracy-robustness-FPR trade-off. Utility-first remains competitive in Macro-F1 but incurs higher staleness, whereas AoI+utility offers a more balanced operating point.

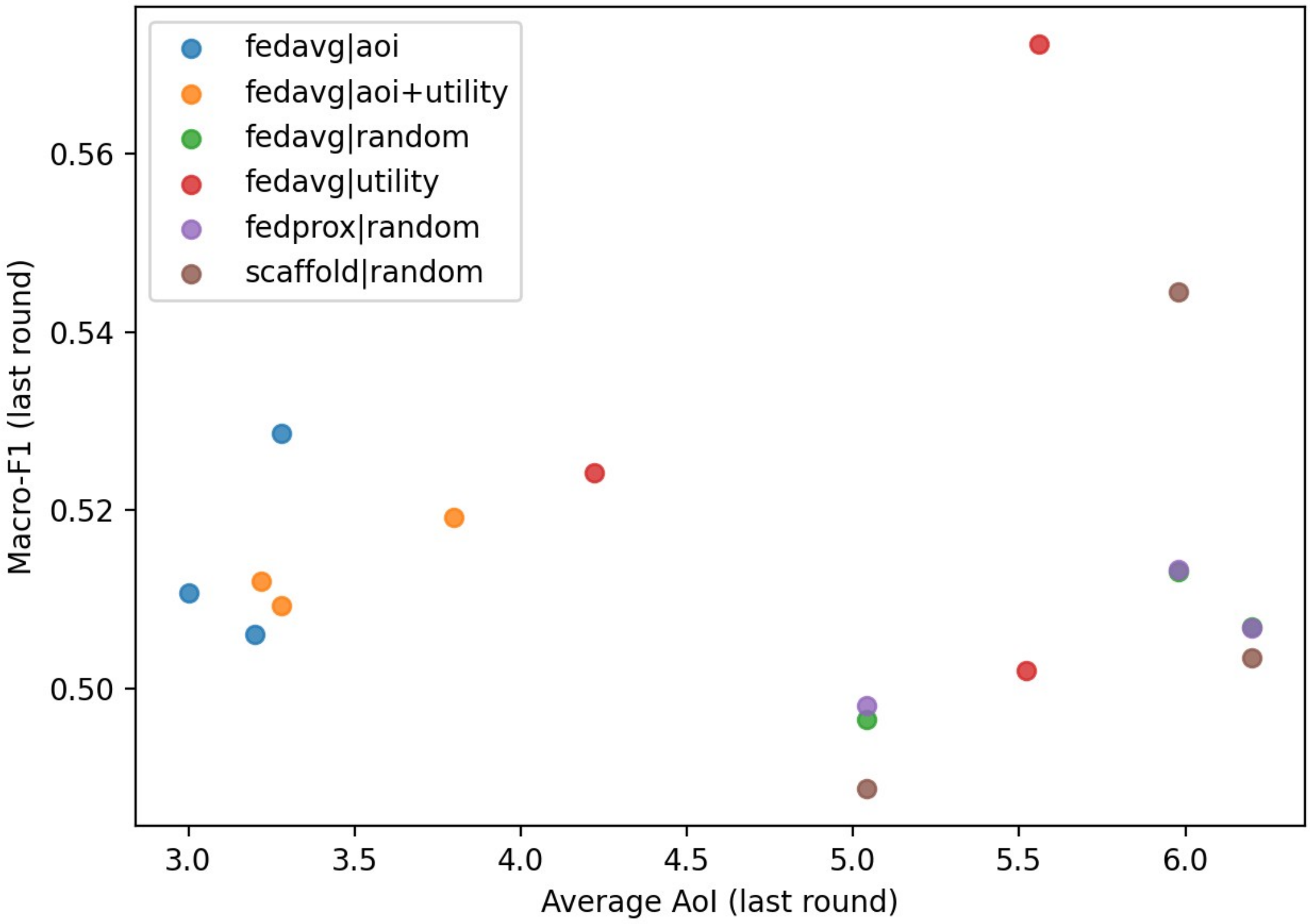

Fig. 2. Accuracy-timeliness trade-off on NSL-KDD under label-flip poisoning with trimmed-mean aggregation (poison_robust).

*Table 3: Sensitivity check of lambda in AoI+utility (synthetic drift, reduced N=30, T=40; 1 seed, illustrative).*

| λ | Macro-F1 | Avg AoI | Peak AoI | Comm (MB) |
|---|---|---|---|---|
| 0.0 | 0.518 | 3.76 | 15.00 | 0.14 |
| 0.3 | 0.532 | 3.08 | 10.00 | 0.15 |
| 0.6 | 0.521 | 3.62 | 12.00 | 0.14 |
| 0.9 | 0.527 | 3.38 | 9.00 | 0.14 |
| 1.0 | 0.524 | 3.38 | 8.00 | 0.14 |

Key findings. (1) Timeliness: AoI-first and AoI+utility consistently reduce Avg AoI by about 39-41% and Peak AoI by about 70% compared with random sampling, across clean and adversarial settings. (2) Detection quality: the hybrid policy is often a practical operating point: it substantially reduces AoI and usually preserves Macro-F1/AUC, but it is not uniformly Pareto-dominant once FPR is included. (3) Robustness: combining trimmed-mean aggregation with AoI+utility maintains low AoI and competitive detection performance under label-flip poisoning. (4) lambda as an explainable knob: Table 3 illustrates that increasing lambda generally reduces staleness (Avg/Peak AoI) with modest impact on Macro-F1, enabling practitioners to tune freshness versus detection performance.

## Discussion, Practical Implications, and Limitations

Our study isolates the effect of client selection using a lightweight classifier. While this clarifies the scheduling question, stronger model classes and additional Byzantine defenses may further improve robustness. For practitioners, the key operational insight is that timeliness can be improved in the FL control plane through client selection and scheduling, rather than through wholesale changes to data pipelines or centralized logging. This is attractive for cloud-edge security analytics, where bandwidth heterogeneity, intermittent availability, and compliance constraints are common. The utility estimate used here deliberately excludes AoI, and the hybrid policy combines utility with freshness explicitly; future work can learn lambda online from workload drift or threat level.

Limitations and future work. We used a lightweight softmax model to isolate the effect of selection; extending to deep temporal models, compression-aware communication, and larger-scale deployments would strengthen external validity. Our poisoning study focuses on label flips; adaptive model poisoning and other Byzantine-robust aggregators (e.g., Krum/median) are promising extensions [10-12]. CIC-IDS2017 results use a reduced mini subset (DDoS and PortScan families), so some metrics (e.g., macro AUC across the full label set) are less informative and omitted. Finally, AoI improves freshness, but freshness alone is not a guarantee of better detection quality or lower false-positive rate; the hybrid policy should therefore be tuned for the target operating environment.

## Conclusion

We presented an AoI-guided client-selection framework for federated intrusion detection in cloud-edge security analytics and evaluated it across multiple IDS datasets and threat scenarios. AoI-aware scheduling substantially improves freshness, and the hybrid AoI+utility policy offers a strong overall trade-off among Macro-F1/AUC, AoI, and robustness while preserving the standard FL communication budget. The central takeaway is systems-oriented: in distributed, privacy-preserving intrusion analytics, better client scheduling can make the detector more timely without requiring heavier models or more communication.